\documentclass[aps,pre,epsf,superscriptaddress,twocolumn,showpacs]{revtex4-1}
\usepackage{titlesec}
\usepackage{graphicx}
\usepackage{epsfig}
\usepackage{dcolumn}
\usepackage{bm}
\usepackage{braket}
\usepackage{amsmath}
\usepackage{amssymb}
\usepackage[titletoc]{appendix}

\begin{document}

\title{Negative-quench-induced excitation dynamics of ultracold bosons \\ in one-dimensional lattices}

\author{S.I. Mistakidis}
\affiliation{Zentrum f\"{u}r Optische Quantentechnologien,
Universit\"{a}t Hamburg, Luruper Chaussee 149, 22761 Hamburg,
Germany}
\author{L. Cao}
\affiliation{Zentrum f\"{u}r Optische Quantentechnologien,
Universit\"{a}t Hamburg, Luruper Chaussee 149, 22761 Hamburg,
Germany}\affiliation{The Hamburg Centre for Ultrafast Imaging,
Universit\"{a}t Hamburg, Luruper Chaussee 149, 22761 Hamburg,
Germany}
\author{P. Schmelcher}
\affiliation{Zentrum f\"{u}r Optische Quantentechnologien,
Universit\"{a}t Hamburg, Luruper Chaussee 149, 22761 Hamburg,
Germany} \affiliation{The Hamburg Centre for Ultrafast Imaging,
Universit\"{a}t Hamburg, Luruper Chaussee 149, 22761 Hamburg,
Germany}

\date{\today}

\begin{abstract}

The nonequilibrium dynamics following a quench of strongly repulsive
bosonic ensembles in one-dimensional finite lattices is investigated
by employing interaction quenches and/or a ramp of the lattice
potential. Both sudden and time-dependent quenches are analyzed in
detail. For the case of interaction quenches we address the
transition from the strong repulsive to the weakly-interacting
regime, suppressing in this manner the heating of the system. The
excitation modes such as the cradle process and the local breathing
mode are examined via local density observables. In particular, the
cradle mode is inherently related to the initial delocalization and,
following a negative interaction quench, can be excited only for
incommensurate setups with filling larger than unity. Alternatively,
a negative quench of the lattice depth which favors the spatial delocalization is used to access the cradle
mode for setups with filling smaller than unity. Our results shed
light on possible schemes to control the cradle and the breathing
modes. Finally, employing the notion of fidelity we study the
dynamical response of the system after a diabatic or adiabatic
parameter modulation for short and long evolution times. The
evolution of the system is obtained numerically using the ab-initio
multi-layer multi-configuration time-dependent Hartree method for
bosons which permits to follow non-equilibrium
dynamics including the corresponding investigation of higher-band effects.\\

\end{abstract}

 \maketitle

\section{Introduction}

The realization of ultracold atomic gases has opened up exciting
possibilities for the study of the non-equilibrium quantum dynamics
of many-body systems \cite{Bloch,Polkovnikov}. The high degree of
tunability and the good isolation from the environment renders
ultracold gases a versatile tool to realize systems far from
equilibrium as they remain coherent for sufficiently long time
scales, allowing us to probe them experimentally \cite{Hung,Ronzheimer,Weidemuller}. In particular, the
dynamical response of a closed quantum system can be investigated
via a sudden change (i.e. a rapid perturbation compared to any other
characteristic time scale of the system) of a Hamiltonian parameter
called 'quantum quench'. Typically, in such a scenario the many-body
system is initially prepared in a characteristic state which is not
an eigenstate of the perturbed Hamiltonian, and the subsequent
time-evolution is explored. In this way, important aspects can be
studied such as the connection between the final and initial states
or the emergence of a steady state \cite{Rigol}.
 Despite recent theoretical advances (see Ref. \cite{Polkovnikov} and references therein),
our understanding of strongly correlated quantum gases after a
quench is far from complete and constitutes an appealing problem of
modern quantum physics
\cite{Cheneau,Natu,Altman,Chen,Haller,Mahmud,Campbell}. 

In a previous work \cite{Mistakidis} following a sudden raise of the interparticle 
repulsion (positive quench) we explored the dynamics of initially weakly interacting 
superfluids. As a consequence a cradle
mode generated by the over-barrier transport of bosons residing in
neighboring wells and caused by the import of energy to the system has been detected.
This mode has further been identified as a two-body intrawell
collision which was dipole-like \cite{Kohn,Bonitz}.  In addition, a
local-breathing mode reminiscent to the usual breathing mode in a
harmonic trap \cite{Abraham,Bauch,Bauch1,Abraham1,Schmitz,Peotta}
has been observed. The occurrence of a resonance between a tunneling
mode and the cradle giving rise to the controlabilibity between the
inter and intrawell dynamics has also been revealed. However, the
above scenario can also give rise to unvoidable heating processes
especially for large quench amplitudes. To overcome this ambiguity i.e. 
minimize the heating \cite{Makotyn} one can start from strong interparticle 
repulsion and quench back to weak interactions called negative interaction quench.
A negative quench may lead to a
drastically different dynamical behaviour as the filling factor
$\nu$ is expected to play a crucial role. Here, an intriguing aspect
would be to explore how the initial spatial configuration of the
system, reflected by the corresponding filling factor, affects the
system dynamics and as a consequence the generation of the emergent
excited modes. This investigation will permit us to gain a deeper
understanding of the on-site excited modes (especially the cradle
mode), the underlying microscopic mechanisms and their
controllability in terms of the tunable parameters of the
Hamiltonian.

In this work a systematic ab-initio analysis of the
non-equilibrium dynamics of strongly repulsive interacting bosons in
one-dimensional (1D) lattices is carried out. To this end, we study from a few-body
perspective the dynamical effects resulting from an abrupt quench or
time-dependent modulation of a Hamiltonian parameter, focussing on
the few-body collective excitations and the control of the dynamics. In
particular, we start from strong repulsive interactions and perform
negative quenches either on the
interparticle repulsion or on the optical lattice depth. This permits us to
unravel the transport properties and the emergent excitation modes
i.e. the local breathing and the cradle processes. Especially,
for the case of a negative interaction quench we demonstrate that
the cradle mode can be excited only for incommensurate setups with filling factor $\nu>1$,
exploiting the initial delocalization. On the other hand,
for filling $\nu<1$ in order to access this mode we use as a
tool a barrier quench, thereby enforcing the over-barrier transport which in turn can generate the cradle mode.
The persistence of the dynamical modes for finite-ramp
rates and long evolution times accessible in recent experiments is also shown. The concept of fidelity is extensively
applied in order to study the response of the quenched system and
the transition from the diabatic to the adiabatic limit. The resulting non-perturbative
dynamics (large quench) is explored using the recently developed numerically exact
multi-layer multi-configuration time-dependent Hartree method for
bosons \cite{Cao,Kronke} (ML-MCTDHB) which reduces in our case of a single species to MCTDHB \cite{Alon,Alon1}.

This article is organized as follows. In Sec.II we
explain the setup, the basic observables and the representation of the wavefunction. Sec.III
is devoted to a detailed study of the non-equilibrium quantum
dynamics for two different quench protocols for incommensurable
setups. We summarize and give an outlook in Sec.IV. Our
computational method ML-MCTDHB is described in the Appendix.

\section{Theoretical Framework}

We consider $N$ identical bosons of mass $m$ trapped within an $n$-site optical lattice along the $x$-direction modeled by the potential
 ${V_\text{tr}}(x) = V_{0} {\sin ^2}\left(\frac{\pi x}{l}\right)$
where $l$ is the distance between successive potential minima,
supplied with hard-wall boundaries at $x =  \pm nl/2$.
Transversally, the bosonic system is trapped by a uniform harmonic
trapping potential with energy spacing $\hbar \omega_\perp$ and
oscillator length ${a_ \bot } = \sqrt{\hbar /m \omega_\bot}$,
yielding an effective 1D coupling strength \cite{Olshanii} $g_{1D} =
\frac{{2{\hbar^2}{a_0}}}{{ma_ \bot ^2}}{\left( {1 - \frac{{\left|
{\zeta (1/2)}\right|{a_0}}}{{\sqrt 2 {a_ \bot }}}} \right)^{ -1}}$
for $s$-wave scattering, ${a_0}$ being the 3D $s$-wave scattering length.
The many-body Hamiltonian then reads
\begin{equation}
\label{eq:1}
H = \sum_{i = 1}^N  -\frac{\hbar ^2}{2m} \frac{\partial^2}{\partial x_i^2} + V_\text{tr}(x_i) + \sum_{i < j} V_\text{int}(x_i - x_j)
\end{equation}
with the short-range contact interaction potential
$V_\text{int}({x_i} - {x_j}) = g_{1D} \delta (x_i - x_j)$ between
bosons located at positions $x_i, x_j$ represented by a Dirac
$\delta$-function. The interaction strength can thereby be tuned by
varying ${a_0}$ via a Feshbach resonance
\cite{Olshanii,Grimm,Duine,Chin} or by altering the extent $a_\perp$
of the transversal confinement \cite{Olshanii,Kim,Giannakeas}. In
the following, for reasons of universality as well as of
computational convenience we shall use dimensionless units. To this
end, the Hamiltonian (1) is rescaled in units of the recoil energy
${E_r} = \frac{{{\hbar ^2k^{2}}}}{{2m}}$. For our simulations we
have used a sufficiently large lattice depth $V_{0}=6.0$ which is of
the order of 3.0 to 4.0 $E_{R}$ (depending on $k$), such that each
well contains at least two localized single-particle Wannier states.
The spatial and temporal coordinates are given in units of ${k^{ -
1}}$ and $\hbar E_r^{ - 1}$, respectively.

A quench is performed by varying, abruptly or slowly, a parameter
$\lambda$ of the system (here, the interaction strength $g_{1D}$ or
the lattice depth $V_{0}$, or generally both) from an initial value
$\lambda_0 = \lambda(t=0)$ to a final value $\lambda_q$ according to
a given scheme $\lambda(t)$. The ground state $\left| {\Psi_0}
\right\rangle$ of the initial Hamiltonian $H_0 = H(\lambda_0)$ then
evolves into $\left|\Psi_\lambda(t) \right\rangle =
U_\lambda(t)\left| {\Psi_0} \right\rangle = \exp(-iH_\lambda
t/\hbar)\left| {\Psi_0} \right\rangle$ at time $t$ under the
$\lambda$-quenched Hamiltonian. The overlap between the time-evolved
states of the system in the presence (via $U_\lambda$) and absence
(via $U_0 = e^{-iH_0t/\hbar}$) of the quench,
\begin{equation}
\label{eq:5}{f_\lambda }(t) = \left\langle
{{\Psi _0}(t)} \right|\left. {{\Psi _\lambda }(t)} \right\rangle,
\end{equation}
yields the fidelity (or Loschmidt echo \cite{Gorin})
\begin{equation}
\label{eq:6} {F_\lambda }(t) = {\left| {{f_\lambda }(t)} \right|^2},
\end{equation}
which provides a time-resolved measure for the effect of the quench
on the system.

Using the ML-MCTDHB method outlined in the Appendix, we obtain the reduced one-body density matrix
\begin{equation}
\label{eq:4} \rho ^{(1)}(x,x';t) = \sum\limits_{a=0}^{M-1}
{{n_a}(t){\varphi _\alpha }(x,t)} \varphi _a^*(x',t)
\end{equation}
in its (diagonal) spectral representation by natural orbitals
${\varphi _\alpha}(x,t)$, where $\alpha=0,1,...,M-1$ and $M$ being
the number of the considered orbitals. The corresponding population
eigenvalues $n_a(t) \in [0,1]$ characterize the fragmentation of the
system \cite{Spekkens,Klaiman,Mueller,Penrose}: If there is only one
macroscopically occupied orbital the system is said to be condensed,
otherwise it is fragmented.

To explore the spatially resolved system dynamics we use the deviation
\begin{equation}
 \delta\rho(x,t) = {\rho}(x,t) - {\left\langle {{\rho(x)}} \right\rangle _T}
\end{equation}
of the one-body density $\rho (x,t) \equiv \rho ^{(1)}(x,x;t)$ from its time-average ${\left\langle {\rho (x)} \right\rangle _T} = \int_0^T {dt \,\rho (x,t)}/T$ over the considered time of propagation $T$.
In this sense, we treat $\delta\rho(x,t)$ as the temporal fluctuation of the density around its ``macroscopic'' component along the lattice.

To incorporate the information of excited bands, we further analyze the dynamics by projecting the many-body wavefunction $\Psi$ to the multiband Wannier number state basis as 
\begin{equation}
\label{eq:7}
\left| \Psi  \right\rangle  = \sum_{N,I} C_{N;I} \left| N_1 \, N_2 \cdots N_n \right\rangle_I,
\end{equation}
where $\{\left| N_1 \, N_2 \cdots N_n \right\rangle_I\}$ is the multiband Wannier number state with $N=\sum_{i}N_i$, and $I$ indexing the energetic (excitation) order \cite{Mistakidis}.
This representation proves convenient for lattice systems when describing intraband and interband processes where the spatial localization of states plays a significant role and
remains valid in the strong interaction regime for a sufficient number of supplied single-particle functions.
Table \ref{table:states} presents the excitation decomposition (the occupation of excited levels in each lattice site) of some number states frequently used in the following analysis.

\renewcommand{\arraystretch}{2.5} %{1.5}
\begin{table}[t]
\caption{
Energetic decomposition of some frequently used number states for $n=3$ lattice sites.
The index $I$ refers to the excitation order and is used as a compact notation instead of the detailed decomposition.
Each element $N^{i}$ in a decomposition refers to the $i$-th energy level (superscript) of $N$ non-interacting bosons in the corresponding site.
}
\centering
\begin{tabular}{l @{\hskip .6in} c @{\hskip .6in} c}
  index $I$ & $ {\left| {2,1,1} \right\rangle _I}$ & ${\left| {1,2,1} \right\rangle _I}$\\
[0.5ex]
%heading
\hline
$I=0$ & $ {| {2^{0},1^{0},1^{0}}\rangle}$  & $ {| {1^{0},2^{0},1^{0}}\rangle}$  \\
$I=1$ & $ {| {1^{0}\otimes 1^{1},1^{0},1^{0}}\rangle}$ & $ {| {1^{1},2^{0},1^{0}}\rangle}$  \\
$I=2$ & $ {| {2^{0},1^{1},1^{0}}\rangle}$ & $ {| {1^{0},1^{0} \otimes 1^{1}  ,1^{0}}\rangle}$\\
$I=3$ & $ {| {2^{0},1^{0},1^{2}}\rangle}$ & $ {| {1^{0},2^{0},1^{1}}\rangle}$ \\
$I=4$ & $ {| {1^{1} \otimes 1^{2},1^{0},1^{0}}\rangle}$ & $ {| {1^{2},2^{0},1^{0}}\rangle}$ \\
$I=5$ & $ {| {2^{0},1^{2},1^{0}}\rangle}$ & $ {| {1^{0},1^{0} \otimes 1^{2} ,1^{0}}\rangle}$ \\
[1ex]
\end{tabular}
\label{table:states}
\end{table}

Note that, the eigenstates can be ordered with respect to the single particle excitation and the spatial occupation of the particles. The eigenstates of the same category 
form an energetical band. Following this categorization we label the eigenstates as ${\left|{\zeta} \right\rangle_{\alpha;I}}$, where $\alpha$, $I$ denote the spatial 
occupation and energetical order respectively \cite{Mistakidis}. For instance $\alpha=1$ refers to a single-pair state, $\alpha=2$ to a two-pair state etc, while $\zeta$ sorts the eigenstates within each category according to 
the eigenenergy.

\section{Quench dynamics}

Before exploring the dynamics, some remarks concerning the ground
states in the lattice for different filling factors $\nu  = N/n$,
where $N$ denotes the particle number and $n$ the number of the
wells, are in order. For the commensurate case ($\nu  = 1,2,...$),
concerning the ground state it is known that for increasing
interparticle interaction one can realize the superfluid to Mott
insulator phase transition \cite{Fisher} which has been addressed
extensively in the past years. On the other hand, for a system with
an incommensurate filling $\nu  \ne 1,2,...$ the main feature is the
existence of a delocalized fraction of particles which forbids the
occurrence of a Mott state. Here, one can distinguish two physical
situations: (a) the case $\nu > 1$ where on-site interaction effects
prevail and (b) $\nu  < 1$ in which the main concern is the
redistribution of the particles over the sites as the interaction
increases. This delocalized phase can also be explained in terms of
the particle hole states using a strong coupling expansion
\cite{Freericks,Freericks1}.

In the present study we consider the quench dynamics for setups with
site occupancy different from unity and therefore exclude the Mott
state physics. We proceed with a short reference to the ground state
and consequently analyze the dynamical process following each quench
protocol.

\begin{figure}[ht]
%\vspace{-15pt}
\centering
\includegraphics[width=0.40\textwidth]{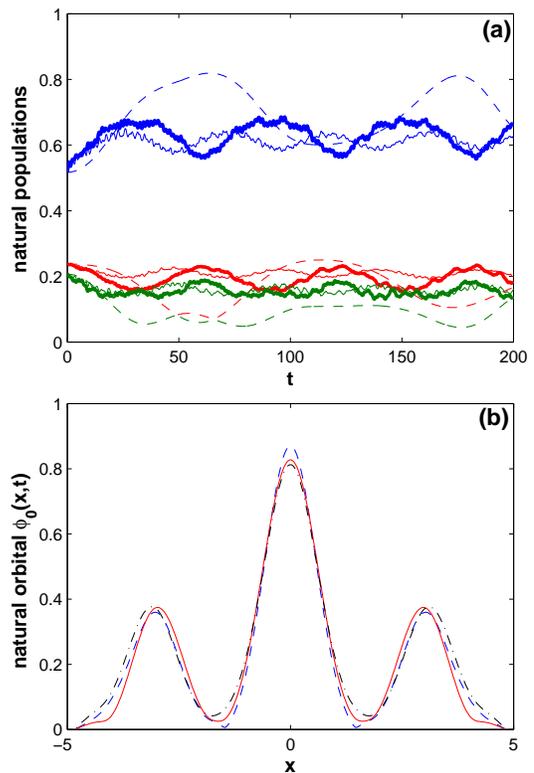}
%  \vspace{1.0cm}
\caption{(Color online) Fragmentation analysis for a system of four
bosons in a triple-well with $g_{in}=5.0$. Shown are (a) the time
evolution of the first three occupations $n_{0}(t)$ (upper panel
blue lines), $n_{1}(t)$ (lower panel red lines) and $n_{2}(t)$ (lower panel green lines), for different
quench amplitudes $\delta{g}=-4.9$ (dashed lines), $\delta{g}=-4.0$
(thick solid lines) and $\delta{g}=-2.5$ (thin solid lines). (b)
Profiles of the lowest natural orbital $\phi_{0}(x,t)$ for a quench amplitude
$\delta{g}=-4.95$ and different time instants during the evolution
$t_{1}=0.9$ (blue dashed), $t_{2}=2.6$ (red solid) and $t_{3}=7.0$ (black dashed-dotted). }
\label{fig:fragmentation}
\end{figure}

\subsection{Quench from strong to weak interactions for filling $\nu>1$}

In this section, we focus on a system consisting of four strongly
interacting bosons in a triple-well, i.e. with filling $\nu>1$. The
initial state before the quench is characterized by the competition
between delocalization and on-site interaction effects. For strong
interparticle repulsion, as we consider here ($g=5.0$), this state
can be interpreted as a fraction $N$ mod $n$ of extra delocalized
particles being on a commensurate background of localized particles.
On the one-body level the on-site populations are quite similar
which can be attributed to the localized background, while their
slight discrepancy is due to the non-uniform distribution of the
extra particle in the first excited band. The latter prevents the
formation of a perfect insulator phase even for strong repulsion.
Our goal is to investigate the dynamical processes following a
negative quench of the interaction strength, thereby approaching the
weakly interacting regime. For an interaction quench protocol the
final Hamiltonian $H_{f}$ can be constructed as a sum of a part
$H_{0}$ which provides the pre-quenched state of the system and an
additional part that denotes the perturbation
\begin{equation}
\label{eq:8}H_{f}({g_f},V) = H_{0}({g_{in}},V) + \frac{{\delta
g}}{{{g_{in}}}}\sum\limits_{k < j} {{V_{int}}({x_k} - {x_j})},
\end{equation}
where $g_{in}$ and $g_{f}$ are the initial and final interaction
strengths respectively, and $\delta{g}=g_{f}-g_{in}$ is the quench
amplitude focussing here on $\delta{g}<0$ and $\left| {\frac{{\delta g}}{{{g_{in}}}}}
\right| \sim 1$.

In the following subsections we first proceed with a brief
fragmentation analysis inspired from the perspective of natural
orbitals. Then, we explain in some detail the response of the system
and investigate each of the emergent normal modes consisting
of a local breathing mode and a dipole-like cradle mode. A study for
the manipulation of the excited modes and their presence for the
case of a finite ramp is also provided.

\subsubsection{Dynamical fragmentation}

In this subsection, we analyze the role of dynamical fragmentation,
i.e. the occurrence of more than one significantly occupied quantum
states during the evolution, with a varying quench amplitude.
Especially, the fragmentation in the non-equilibrium dynamics of
trapped finite systems is known to depend strongly on the particle
number \cite{Mueller,Sakmann}, the interaction strength and the
evolution time. The spectral decomposition of the one-body reduced
density matrix offers a measure of fragmentation via the populations
$n_{a}(t)$ of the natural orbitals $\phi_{a}(t)$ (see eq.(4)). In
particular, a non-fragmented (condensed) state requires the
occupation of $n_{0}(t)$ to be close to unity \cite{Penrose}.
\begin{figure}[t]
%\vspace{-15pt}
\centering
\includegraphics[width=0.40\textwidth]{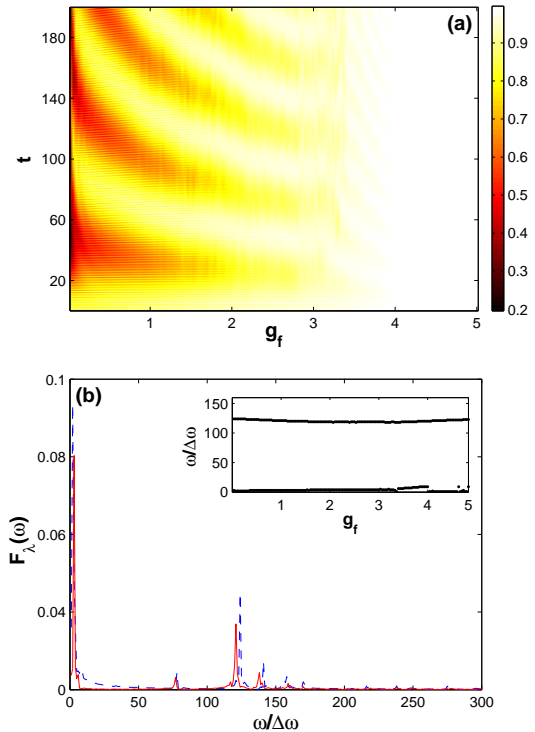}
%  \vspace{1.0cm}
\caption{(Color online) (a) Fidelity evolution following negative interaction quenches for $g_{in}=5.0$.
(b) The frequency spectrum of the fidelity for
$g_{f}=0.6$ (blue dashed) and $g_{f}=1.0$ (red solid),
which indicates the tunneling modes. The inset shows the dependence of each tunneling branch with respect to the final interaction strength after the quench.
We incorporate 150 quenches in the
range of $0<g_{f}<5.0$. The
frequency units are normalized as $\omega/\Delta\omega$, with $\Delta\omega=2\pi/T$
and $T$ being the considered evolution time.}
\end{figure}

Figure 1(a) shows the evolution of the natural
populations of the three highest occupied natural orbitals for different quench amplitudes. The population of the
first orbital $n_{0}(t)$ is always significantly below unity which
confirms the fragmentation process, while the three most occupied natural orbitals
add up to more than $90\%$ of the population. Focussing on the first
orbital we note that the temporal average of the fragmentation reduces as the quench amplitude 
increases and vice versa. Especially, for final interactions close to a
non-interacting state we observe a tendency for a non-fragmented
state at least for certain time periods. This constitutes a major
difference between a negative and a positive interaction quench
scenario. In the latter case the fragmentation process is enhanced
for larger quench amplitudes which can be attributed to the
consequent raise of the interparticle repulsion during the process.
However, here we face the inverse behaviour because in the initial
strongly interacting state the interparticle repulsion is already
significant and tends to be reduced after the quench. Moreover, the second and third orbitals
take on a compensatory role to the first, e.g. in the time periods
where $n_{0}(t)$ is enhanced $n_{1}(t)$ and $n_{2}(t)$ are reduced.
Finally, note that for smaller quenches the latter populations
possess smaller amplitude oscillations whereas strong quenches
introduce large amplitude variations of the populations.

Figure 1(b) illustrates the response of the first natural orbital
$\phi_{0}(x,t)$ at different time-instants during the evolution
after a quench to $g_{f}=0.05$. As it can be seen $\phi_{0}(x,t)$
exhibits spatial oscillations in the outer wells and an on-site
broadening in the middle well which accounts for interaction
effects. Another important remark is that the band-structure is
effectively reflected by the population of the natural orbitals,
i.e. the orbitals $\phi_{0}(x)$, $\phi_{1}(x)$ and $\phi_{2}(x)$
correspond to the effective first single-particle band, orbitals $\phi_{3}(x)$,
$\phi_{4}(x)$ and $\phi_{5}(x)$ to the second band etc. Thus, the
lowest orbital $\phi_{0}(x,t)$ follows quite well the evolution of
the quenched one-body density.

\subsubsection{Dynamical response and transport properties}

To investigate the dynamical response of the system we use the
above-discussed fidelity $F_{\lambda}(t)$ (see eq.(3)). This
quantity is shown in Figure 2(a) as a function of the final
interaction strength and the time. We mainly note the appearance of
two different regions as a function of the quench amplitude. The
first one corresponds to quenches from a strong repulsive state with
$g_{in}=5.0$ to intermediate interactions where $3.4<g_{f}<5.0$.
Here, the overlap during the dynamics is rather large with minimal
percentage up to $85\%$ and therefore the system is quite
insensitive to the quench. In the second region where the final
state belongs to weak or even to the non-interacting regime, i.e.
$0<g_{f}<3.4$, we observe the formation of an oscillatory pattern in
the fidelity evolution. This pattern indicates the sensitivity of
the system to these type of quenches meaning that the system can be
driven far from the initial state, while the minimal overlap for the
extreme case of $g_{f}\to0$ can even be of the order of $20\%$. The
emergence of the above regions is universal in the system in the
sense that it weakly depends on the height of the barrier. Thus, for
an increasing barrier height the second region (larger quenches)
will become narrower due to the larger potential energy which
inhibits a possible departure of the system from the initial state.

\begin{figure*}[t]
%\vspace{-15pt}
\centering
\includegraphics[width=0.90\textwidth]{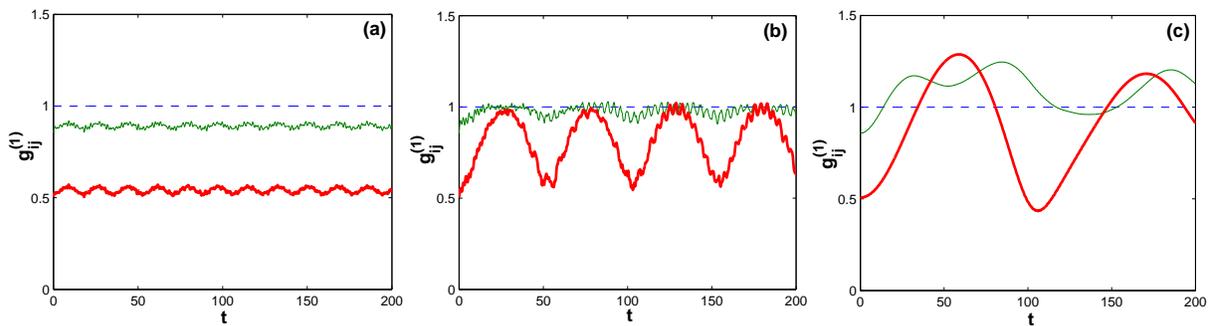}
%  \vspace{1.0cm}
\caption{(Color online) The time evolution of the one-body correlation function $g^{(1)}_{ij}$ after various negative interaction quenches for $g_{in}=5.0$. Shown are
different components of the correlation function $g^{(1)}_{ij}$ with respect to the left well $g^{(1)}_{LL}$ (blue dashed line), $g^{(1)}_{LM}$ (green thin solid line) and
$g^{(1)}_{LR}$ (red thick solid line) for final interactions (a) $g_{f}=3.8$, (b) $g_{f}=1.6$ and
(c) $g_{f}=0.05$.}
\end{figure*}
In order to identify the tunneling modes participating in the
dynamics we use as a measure the spectrum of the fidelity
${F_\lambda }(\omega ) = \frac{1}{\pi }\int {dt{F_\lambda
}(t){e^{i\omega t}}}$. Indeed, Figure 2(b) shows ${F_\lambda
}(\omega )$ for different final interactions where we observe two
dominant tunneling peaks. To proceed with a more quantitative
description of the tunneling dynamics we shall expand the
wavefunction in terms of the number states. To this end, let $\left|
{\Psi (0)} \right\rangle  = \sum\limits_{\zeta ;\alpha ;I} {C_\zeta
^{\alpha ;I}{{\left| \zeta \right\rangle }_{\alpha ;I}}}$ be the
initial wavefunction in terms of the eigenstates ${{{\left| \zeta
\right\rangle }_{\alpha ;I}}}$ of the final Hamiltonian \cite{Mistakidis}. Then, the
expansion of the fidelity reads
\begin{equation}
\begin{split}
\label{eq:9}{\left| {\left\langle {\Psi (0)} \right|\left. {\Psi
(t)} \right\rangle } \right|^2} = \sum\limits_{\zeta_{1};\alpha;I }
{{\left| {{C_{\zeta_{1}}^{\alpha;I}}} \right|}^4} +
\sum\limits_{\zeta_{1},\zeta_{2};\alpha ,\beta;I } {{\left|
{{C_{\zeta_{1}}^{\alpha;I}}} \right|}^2}\\\times{{\left|
{{C_{\zeta_{2}}^{\beta;I}}} \right|}^2}\cos ({\epsilon
_{\zeta_{1}}^{\alpha;I}} - {\epsilon _{\zeta_{2} }^{\beta;I}})t,
\end{split}
\end{equation}
where the second term contains the separate contributions from each
tunneling branch. The indices $\alpha$, $\beta $ indicate a
particular set of number states, ${\zeta _i}$ is the
intrinsic index within each set, $I$ denotes the respective
energetical level and ${\epsilon _{\zeta_{i} }^{\alpha;I}}$ refers to the eigenfrequency
of a particular eigenstate. In
particular, the first peak at frequency ${\omega _1} \approx
3\Delta\omega$ (with $\Delta\omega=2\pi/T$ and $T$ denotes the
propagation time) corresponds to the energy difference
$\Delta\epsilon$ within the energetically lowest states of the
single pair mode. Therefore the process corresponds to an intraband
tunneling, e.g. from the state ${\left| {1,2,1} \right\rangle _0}$
to ${\left| {2,1,1} \right\rangle _0}$ etc. However, the second peak
located at $\omega_2 \approx 125\Delta\omega$ refers to an interband
transition between the states ${\left| {1,2,1} \right\rangle _2}$
and ${\left| {1,2,1} \right\rangle _0}$, which reflects the initial
strongly correlated state. In the inset we present the
$\delta{g}$-dependence of the location of the aforementioned peaks.
As it can be seen the two branches are mainly steady as a function
of the interaction quench, their frequencies are constrained in a
narrow-band, while their amplitude (see main Figure) reduces
significantly for weak quenches.

In the course of the investigation of the tunneling dynamics one
fundamental question that has to be addressed is how
correlations propagate \cite{Bravyi} in the quenched system. 
Here, in order to distinguish genuine interwell correlations from density 
oscilation effects we explore the response of the normalized single particle correlations ${g^{(1)}_{ij}}(t) =
\left\langle \Psi \right| {b_i}b_j^\dag\left| \Psi
\right\rangle/\sqrt{\left\langle \Psi \right| {b_i}b_i^\dag\left| \Psi
\right\rangle \left\langle \Psi \right| {b_j}b_j^\dag\left| \Psi
\right\rangle} $ \cite{Sakmann1}. $b_i^\dag$ denotes the creation
operator of a particle at the $i-th$
well, while the diagonal elements $g^{(1)}_{ii}=1$ by definition. An important property of this 
function is that for ${g^{(1)}_{ij}}>1$ ($<1$) the corresponding detection probabilities at positions $i$ and $j$ are 
correlated (anticorrelated), while the case ${g^{(1)}_{ij}}=1$ is referred to as fully first order coherent.
Figures 3(a)-(c) illustrate the time evolution for
different components of the one-body correlations for
various negative interaction quenches. As expected the diagonal terms correspond to a straight line at unity for all quench amplitudes. 
 The non-diagonal terms $g^{(1)}_{ij}$, $i\neq j$ exhibit a non-vanishing oscillatory
behaviour, while for increasing
quench amplitude a substantial built-up of
correlations is observed. In particular, approaching the non-interacting limit ${g^{(1)}_{LM}}(t)>1$ 
for most of the time, whereas ${g^{(1)}_{LR}}(t)$ oscillates around unity indicating a transition from an anticorrelated to a correlated situation.

\subsubsection{The local breathing mode}

The breathing mode can be used in order to
measure some key quantities of a trapped system such as its kinetic
and interaction energy or the coupling strength \cite{Abraham,Abraham,Bauch,Bauch1}. It refers to
an expansion and contraction of the bosonic cloud and can be excited
either via a variation of the interparticle interaction or a modulation of
the frequency of the trapping potential.

In a similar manner, our quenched system exhibits local breathing
oscillations which are most prominent in the subsystem corresponding
to the middle well. To detect the frequencies of this normal mode we
examine the variance of the coordinate of the center of mass for a
particular well. The center of mass for the $i-th$ well is defined
as $X_{CM}^{(i)} =\int_{d_i}^{d'_{i}}dx \left( {x - x_0^{(i)}}
\right){\rho _i}(x)/\int_{d_i}^{d'_{i}}dx{\rho _i}(x)$. The index
$i=R,M,L$ corresponds to the right, middle and left well
respectively, while ${x_0^{(i)}}$ refers to the central point of the
corresponding well. The limits of the wells are denoted by ${d_i}$
and ${d'_i}$, whereas ${\rho _{i}}(x)$ is the respective single
particle density. For the identification of the breathing process we
define the operator of the second moment $\sigma _M^2(t) =
\left\langle {{{\Psi|\left( {x - X_{CM}^{(i)}} \right)}^2}}
|\Psi\right\rangle $. The latter serves as a measure for the
instantaneous spreading of the cloud in the $i-th$ well and can also
be used experimentally in order to probe the expansion velocity of a
quenched condensate \cite{Ronzheimer}. Then, if we connect the
initial wavefunction with the eigenstates ${\left| \zeta
\right\rangle _{\alpha ;I}}$ of the final Hamiltonian $H_{f}$, we
obtain
\begin{equation}
\begin{split}
\label{eq:10}\begin{array}{l}
\sigma _M^2(t) = \sum\limits_{\alpha ;\zeta_{1};I} {{{\left| {C_{\zeta_{1}}^{a;I}} \right|}^2}{}_{\alpha ;I}\left\langle \zeta_{1} \right|} {\left( {x - X_{CM}^{(i)}} \right)^2}{\left| \zeta_{1} \right\rangle _{\alpha ;I}} \\
 + 2\sum\limits_{\zeta_{1} \ne \zeta_{2}} {{\mathop{\rm Re}\nolimits} \left( {C_{\zeta_{1}}^{\beta ;I*}C_{\zeta_{2}}^{\alpha ;I}} \right){}_{\beta ;I}\left\langle \zeta_{1} \right|} {\left( {x - X_{CM}^{(i)}} \right)^2}{\left| \zeta_{2} \right\rangle _{\alpha ;I}}\\~~~~~~~~~~~~~~~~~~~~~~~~~~~~~\times\cos \left( {\omega _{\zeta_{1}}^{\beta ;I} - \omega _{\zeta_{2}}^{\alpha ;I}} \right)t.
\end{array}
\end{split}
\end{equation}
To identify the frequencies of the local breathing mode Figure 4
shows the frequency spectrum of the second moment $\sigma
_M^2(\omega ) = \frac{1}{\pi}\int{dt\sigma_M^2(t){e^{i\omega t}}}$,
which refers to the middle well, for different quench amplitudes.
Three main peaks can be observed. The lowest of these three peaks
refers to a tunneling mode being identified from the energy
difference within the energetically lowest states of the single pair
mode. The appearance of this peak in the spectrum is due to the fact
that the tunneling can induce a change in the width of the local
wavepacket. The second and third peaks refer to interband processes
and are related to higher-band transitions. In particular, the
second peak is located at $\omega_{2} \approx 125\Delta\omega$ and
refers to a transition from ${\left| {1,2,1} \right\rangle _0}$ to
${\left| {1,2,1} \right\rangle _2}$, whereas the third one with
frequency $\omega_{3} \approx 170\Delta\omega$ corresponds to a
transition from ${\left| {1,2,1} \right\rangle _0}$ to ${\left|
{1,2,1} \right\rangle _5}$. To illustrate the dependence of the
above three peaks on the interaction quench we show in the inset the
evolution of the location of each peak with respect to the final
interaction strength $g_{f}$ after the quench. We observe that the
branches are more sensitive for a quench to $2.0<g_{f}<4.0$,
otherwise they are mainly constant.
\begin{figure}[t]
%\vspace{-15pt}
\centering
\includegraphics[width=0.80\columnwidth]{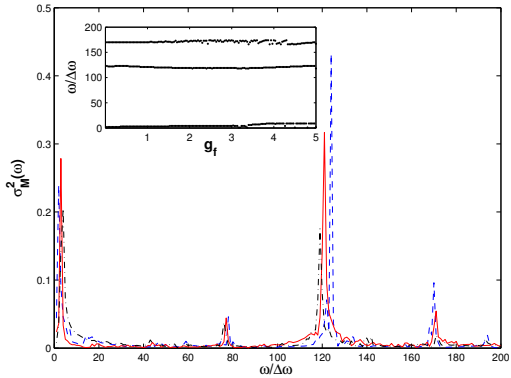}
%  \vspace{1.0cm}
\caption{(Color online) Fourier spectrum of the second moment $\sigma _M^2(\omega )$ for the local breathing mode
for different quench amplitudes. The initial state corresponds to
$g_{in}=5.0$  and the final interactions are $g_{f}=0.15$ (blue dashed), $g_{f}=0.8$ (red solid),
and $g_{f}=1.7$ (black dashed-dotted). In the inset we show the $\delta{g}$-dependence of each breathing branch, where we incorporate 150 quenches in the
range of $0<g_{f}<5.0$. Note that the frequency units are
normalized with respect to $\omega/\Delta\omega$,
where $\Delta\omega=2\pi/T$ and $T$ is the respective propagation time.}
\end{figure}

\subsubsection{The cradle mode}

This mode refers to a dipole-like oscillation generated via an
over-barrier transport due to the initially delocalized state
between neighboring wells. In the present case it is induced by
an interaction quench. From a one-body perspective the cradle mode
is demonstrated by the inner well dynamics of the one-body density
fluctuations $\delta\rho(x,t)$. Figure 5 shows the evolution of the
system through the relative density after a sudden negative
interaction quench from $g_{in}=5.0$ to $g_{f}=0.07$. The emergence
of the cradle mode in the outer wells manifested as a dipole-like
oscillation and the local breathing in the central well as a
contraction and expansion dynamics is observed.

The initial spatial configuration due to the strong interparticle
interaction corresponds to one localized boson in each well and one
delocalized (over the three wells) energetically close to the
barrier. In turn, the negative change in the interaction strength
yields a high probability for the delocalized particle to overcome
the barrier (over-barrier transport) and move to a neighboring well,
where it performs a collision with the initially localized particle.
This results in the cradle-like mode inside the respective
neighboring site and refers to a localized wave-packet oscillation \cite{Mistakidis}.
Note that the cradle is inherently related to the initial
delocalization and after a negative interaction quench of a strongly
correlated system can be excited only for incommensurate systems
with filling factor $\nu>1$. For other fillings it disappears and
the consequent dynamics is dominated by the interwell tunneling.

In the following, in order to quantitatively examine the inner-well
dynamics we proceed with a local density analysis. For that purpose
we divide a particular well from the center into two equal sections
with ${\rho _{a,1}}(t)$ and ${\rho _{a,2}}(t)$ being the respective
integrated densities of the left and right parts during the
evolution. The index $a = L,M,R$ stands for the left, middle and
right well respectively. In this manner, a measure of the intrawell
asymmetry which captures the cradle motion is the quantity $\Delta
{\rho _a}(t) = {\rho _{a,1}}(t) - {\rho _{a,2}}(t)$.  Figure 6(a)
shows the frequency spectrum of the above quantity for the left
well, i.e. $\Delta {\rho _L}(\omega ) = \frac{1}{\pi }\int {dt}
\Delta {\rho _L}(t){e^{i\omega t}}$ for various negative interaction
quenches. From the spectrum we can identify two dominant peaks
located at the positions ${\omega _2} \approx 79\Delta \omega$ and
${\omega _3} \approx 125\Delta \omega$. These two frequency branches
correspond to the cradle mode. In addition, we observe a
low-frequency peak related to the interwell tunneling at frequency
${\omega _1} \approx 3\Delta \omega$. The inset shows the
$\delta{g}$-dependence of the above three frequency peaks. The
location of each branch remains essentially independent of the
strength of the interaction quench and it is therefore constrained
to a corresponding narrow-band.
\begin{figure*}[t]
%\vspace{-15pt}
\centering
\includegraphics[width=0.80\textwidth]{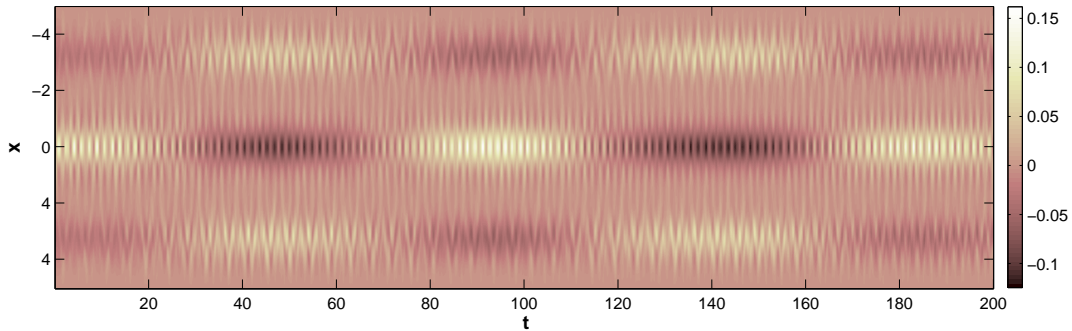}
%  \vspace{1.0cm}
\caption{(Color online) Space-time evolution of the fluctuations
$\delta\rho(x,t)$ after a sudden negative quench of the
inter-particle repulsion from $g_{in}=5.0$ to $g_{f}=0.07$, thereby
approaching the non-interacting limit. We observe the cradle mode in
the left and right wells, the local breathing mode in the middle
well and the interwell tunneling during the evolution.}
\end{figure*}

To gain a deeper understanding of the cradle mode we again refer to
a number state analysis and expand the initial state of the system
in terms of the eigenstates of the final Hamiltonian as $\left|
{\Psi (0)} \right\rangle = \sum\limits_{\zeta ;\alpha ;I} {C_\zeta
^{\alpha ;I}{{\left| \zeta \right\rangle }_{\alpha ;I}}}$. Then, the
expectation value of the asymmetry operator reads
\begin{equation}
\begin{split}
\label{eq:13}\begin{array}{l}\left\langle \Psi  \right|\Delta
\widehat \rho(t) \left| \Psi  \right\rangle  =
\sum\limits_{\zeta_{1};\alpha;I}  {{{\left|
{{C_{\zeta_{1}}^{\alpha;I} }} \right|}^2}{}_{I;\alpha}\left\langle
\zeta_{1} \right|} \Delta \widehat \rho \left| \zeta_{1}
\right\rangle_{\alpha;I} \\
~~~~~~~~+ 2\sum\limits_{\zeta_{1}  \ne \zeta_{2} } {{\mathop{\rm
Re}\nolimits} \left( {C_{\zeta_{1}}^{\alpha;I
*}{C_{\zeta_{2}}^{\beta;I} }} \right){}_{I;\alpha}\left\langle \zeta_{1}  \right|}
\Delta \widehat \rho \left| \zeta_{2}
\right\rangle_{\beta;I}\\~~~~~~~~~~~~~~~~~~~~~~~~~~~~~~\times \cos
\left[ {\left( {{\omega _{\zeta_{1}}^{\alpha;I} } - {\omega
_{\zeta_{2}}^{\beta;I} }} \right)t} \right].\end{array} \end{split}
\end{equation}
Here the terms of the second sum in the above expression which
demonstrate an oscillatory behaviour describe the cradle mode.
Therefore, we need to detect the eigenstates (${{{\left| \zeta
\right\rangle }_{\alpha ;I}}}$) of the dominant oscillation terms,
i.e. ${}_{\alpha ;I}\left\langle \zeta  \right|\Delta \widehat \rho
{\left| \zeta  \right\rangle _{\beta ;I}} \ne 0$. A direct numerical
analysis indicates that the respective eigenstates are ${\left|
\zeta  \right\rangle _{1;0}}$, ${\left| \zeta  \right\rangle
_{1;1}}$, ${\left| \zeta  \right\rangle _{1;2}}$, whereas the
corresponding significantly contributing number states are ${\left|
{2,1,1} \right\rangle _0}$, ${\left| {2,1,1} \right\rangle _1}$ and
${\left| {2,1,1} \right\rangle _4}$ due to the fact that the
corresponding oscillation frequency matches the energy difference
between these eigenstates.
\begin{figure}[t]
%\vspace{-15pt}
        \centering
                \includegraphics[width=0.50\textwidth]{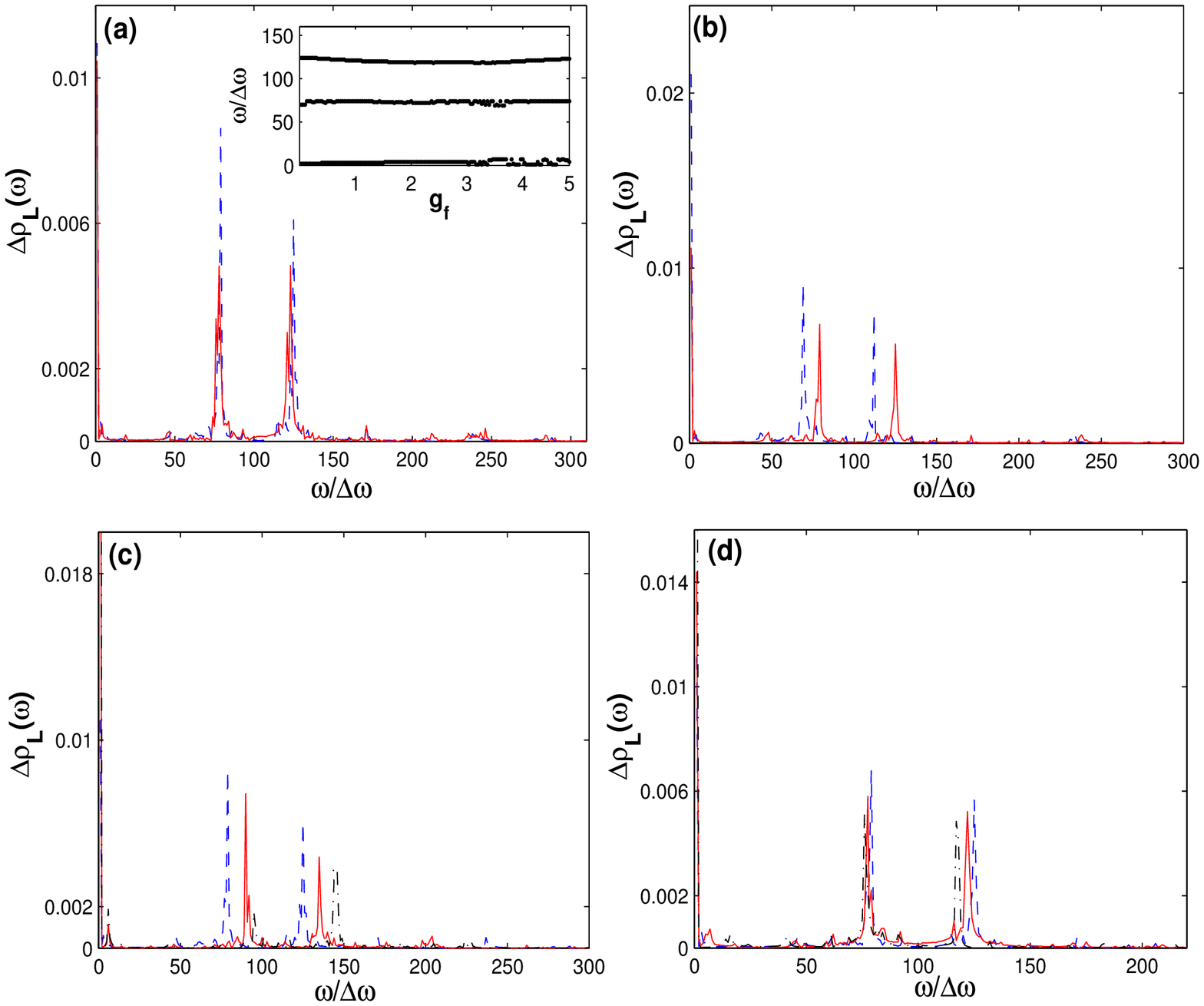}
              %  \vspace{1.0cm}
                \caption{(Color online) The frequency spectrum of the intrawell asymmetry $\Delta\rho_{L}(\omega)$.
                (a) The final state of the system is obtained after a sudden negative interaction quench from
                $g_{in}=5.0$ to $g_{f}=0.1$ (blue dashed) and $g_{f}=0.45$ (red solid). The inset shows the evolution of each peak that refers to the cradle as a function
                of the quench amplitude (we incorporate 150 quenches in the
                range $0<g_{f}<5.0$). (b) The spectrum $\Delta\rho_{L}(\omega)$ for the
                same quench amplitude, $\delta{g}=-4.95$, and different barrier heights $V_{0}=5.5$ (red solid) and $V_{0}=3.5$ (blue dashed).
                (c) Sudden quench to $g_{f}=0.4$ and the hard-wall boundaries located at $x_{\sigma}=\pm3\pi/2$ (blue dashed), $x_{\sigma}=\pm5\pi/4$ (red solid)
                and $x_{\sigma}=\pm11\pi/8$ (black dashed-dotted). (d) It is illustrated the spectrum of $\Delta\rho_{L}(\omega)$ for an imposed harmonic trap $V_{harm}=0$ (blue dashed), $V_{harm}=0.02x^2$ (red solid) and $V_{harm}=0.05x^2$ (black dashed-dotted) on top of the lattice.
                Finally, note that in each case we use normalized frequency units $\omega/\Delta\omega$, with $\Delta\omega=2\pi/T$ and $T$ being
                the respective evolution time.}
\end{figure}

Let us now, investigate possible control protocols of the cradle
mode via a modulation of its frequency by means of a varying
potential parameter or via an external forcing. An efficient way to
manipulate the frequency is to tune the height of the potential
barriers. In this way, the cradle frequency can be reduced using a
more shallow lattice (thereby making the excitation of the cradle
mode more easy). Indeed, within the harmonic approximation it can be
easily shown that the effective frequencies for two lattices with
different potential depths $V_{0;1}$ and $V_{0;2}$, respectively,
obey $\omega_{eff;1}=(V_{0;1}/V_{0;2})^{1/4}\omega_{eff;2}$. This
situation is illustrated in Figure 6(b) where the frequency spectrum
of the inner-well asymmetry with the same quench amplitude but
different barrier heights $V_{0}=5.5$ (red solid) and $V_{0}=3.5$
(blue dashed) is shown. We observe a negative shift of each
frequency peak for a decreasing lattice depth which confirms our
previous arguments. Alternatively, a similar manipulation of the
cradle frequency can be achieved by comparing lattices with the same
height of the potential barrier but different frequencies. Then, the
respective effective frequencies are related via
$\omega_{eff;1}=(l_{2}/l_{1})^{1/2}\omega_{eff;2}$, where $l$ is the
distance between two successive potential minima.

In a similar manner, one can pose the question how the cradle mode
frequency depends on $g_{in}$ for fixed $g_{f}$. According to our
simulations (omitted here for brevity) each peak remains essentially
unchanged, indicating that the system does not keep any memory from
the particular strongly correlated initial microscopic
configuration.

A further question is to ask for the impact of the boundary
conditions. Hence, we assume a fixed height for the barrier but
changing the position of the hard wall boundary conditions. Then, we
expect that as the wall is closer to the center of the right or left
well the cradle would be more enhanced because effectively the
frequency of the local harmonic oscillator is larger and so the
period of the cradle reduces. Indeed, Figure 6(c) illustrates for
the same quench amplitude the Fourier spectrum of the intrawell
asymmetry $\Delta\rho_{L}(\omega)$ imposing the hard-wall boundaries
at different positions, namely at $x_{\sigma}=\pm3\pi/2$ (blue
dashed), $x_{\sigma}=\pm5\pi/4$ (red solid) and
$x_{\sigma}=\pm11\pi/8$ (black dashed-dotted). The frequency peaks
of the cradle mode are shifted by a positive value for a closer to
the center hard-wall. As a final attempt we impose a harmonic trap
on top of the triple well, which increases the potential energy of
the edge wells. Then the on-site energy of the Wannier states at the
edges becomes larger than that (of the same degree of energetical
excitation) in the central well. This in turn renders the
initialization of the cradle mode more difficult, and for strong
superimposed harmonic traps its excitation for a fixed quench
amplitude becomes impossible. Accordingly, Figure 6(d)
shows a scenario with the same quench amplitude but different
superimposed harmonic traps. We observe negative variations and a
reduction of the intensity of each peak for a stronger harmonic
trap, thereby confirming our above discussion.
\begin{figure}[t]
%\vspace{-15pt}
\centering
\includegraphics[width=0.50\textwidth]{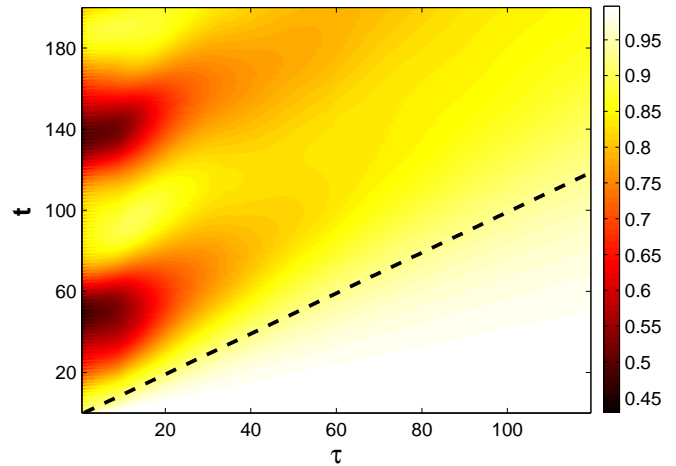}
%  \vspace{1.0cm}
\caption{(Color online) Fidelity evolution for $g_{in}=5.0$ and
$g_{f}=0.1$ as a function of the ramp-rate $\tau$ measured in units
of the Heisenberg time $\tau_{H}$ (see text). The black dotted line
correspond to the situation with $t=\tau$.}
\end{figure}

In the next subsection we explore the excitation modes induced by a
time-dependent modulation of the interaction strength and
establish their presence also for this case.

\subsubsection{Finite ramping}

The present subsection is devoted to the study of the dynamics
induced by time-dependent interaction quenches with a finite
ramp-rate. In particular, we attempt to investigate quenches with
the same amplitude but evolving on different time-scales, in order
to gain a further insight into the dynamical response of the system
with relevance to the experimentally occurring time-scales. To this
end, let us adopt a time-dependent quench scenario of the form
\begin{equation}
\label{eq:15}g(t;\tau) =g_{in} + (g_{f}-g_{in})\tanh (t/\tau ).
\end{equation}
Here $g_{in}$, $g_{f}$ are the interaction strength for the initial
and final state respectively, whereas $\tau$ denotes the finite
ramp-rate of the performed quench. Focussing now on a strong
non-equilibrium post-quench state with $g_{f}=0.1$, Figure 7 shows
the dynamical crossover, for finite evolution times, from an abrupt
to an adiabatic interaction modulation for increasing ramp-rates
$\tau$. To interpret the resulting behaviour on a relevant
time-scale we define the Heisenberg time
$\tau_{H}\sim1/\Delta\epsilon(\delta{g})$, where
$\Delta\epsilon(\delta{g})=\epsilon({g_{in}})-\epsilon({g_{f}})$
refers to the energy difference between the ground state of the
system before and after a sudden interaction quench. As it is shown
for times $t<\tau$ (region under the black dotted line in Figure 7)
the system essentially remains in the initial ground state of the
unperturbed Hamiltonian. On the contrary, in the region with
$t>\tau$, which spreads for decreasing $\tau$ (thereby approaching
the sudden quench), the system starts to significantly depart from
the initial state. Remarkably enough for $\tau<30\tau_{H}$ we
observe the appearance of black lobes (overlap of the order of
$40\%$) during the evolution which indicate the persistence of the
excitation modes in this region. For $\tau>30\tau_{H}$ we have a
transition to a smoother dynamical departure of the system from the
initial state and as a consequence the elimination of the excitation
modes. In particular, for $\tau>85\tau_{H}$ the Hamiltonian changes
sufficiently slowly, i.e. the system tends to remain in the
instantaneous ground state and therefore the modulation is almost
adiabatic for the whole evolution time. For a smaller quench the
adiabatic regime can be reached for sufficiently smaller time-scales
due to the reduced impact of the quench to the system. These
statements are also valid for a linear quench protocol of the form
$g(t;\tau)=g_{in}+(g_{f}-g_{in})t/\tau$ for $t\leq\tau$, and
$g(t;\tau)=g_{f}$ for $t>\tau$.
\begin{figure*}[t]
%\vspace{-15pt}
        \centering
                \includegraphics[width=0.80\textwidth]{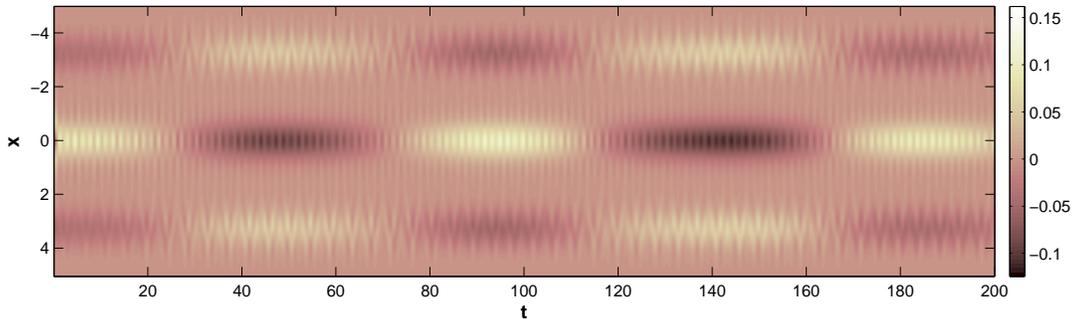}
              %  \vspace{1.0cm}
                \caption{(Color online) The fluctuations $\delta\rho(x,t)$ of the one-body density caused by a negative time-dependent quench
                of the inter-particle repulsion to $g_{f}=0.07$ ($g_{in}=5.0$) with a finite ramp-rate
                $\tau=0.8\tau_{H}$. For a direct comparison the quench parameters, here, have been chosen
                similar with Figure 5, which refers to
                the respective sudden quench scenario. We observe that the cradle mode
                in the left and right wells, the local breathing mode in the middle well and the interwell tunneling during
                the evolution persist.}
\end{figure*}

As a next step we study the effect of the time-dependent interaction
quench on the excitation modes, i.e. the breathing and cradle
processes. To give further insight in the overall non-equilibrium
process Figure 8 illustrates the evolution of $\delta\rho(x,t)$, for
the same relevant interactions as in Figure 5 where we considered a
sudden quench, implementing now the time-dependent scenario of
equation (11) with a finite rate $\tau=0.8\tau_{H}$. The
above-discussed modes still persist but with reduced intensity which
is larger when the quench is faster.

According to this let us investigate how one can manipulate the
local breathing mode via the quench  rate $\tau$. Figure 9(a) shows
the frequency spectrum of the local breathing mode obtained for the
same amplitude $\delta{g}=-4.9$ and different quench rates $\tau$.
As it can be seen the position of each peak remains the same but its
intensity decreases significantly for larger rates. To further probe
the position of each branch with respect to the quench rate $\tau$
we present in the inset the $\tau$-dependence of each peak (without
taking into account its intensity). It is obvious that each branch
is quite insensitive to the interaction quench while in terms of its
intensity (Figure 9(a)), one can infer that by considering larger rates
can gradually obliterate each frequency branch, i.e. for a faster
quench the spectrum is more rich. Especially, one finds that for
$\tau>30\tau_{H}$ this mode can essentially be eliminated which
means that the intensity of each peak is negligible (in our case
$\leq10^{-5}$).
\begin{figure}[t]
%\vspace{-15pt}
        \centering
                \includegraphics[width=0.40\textwidth]{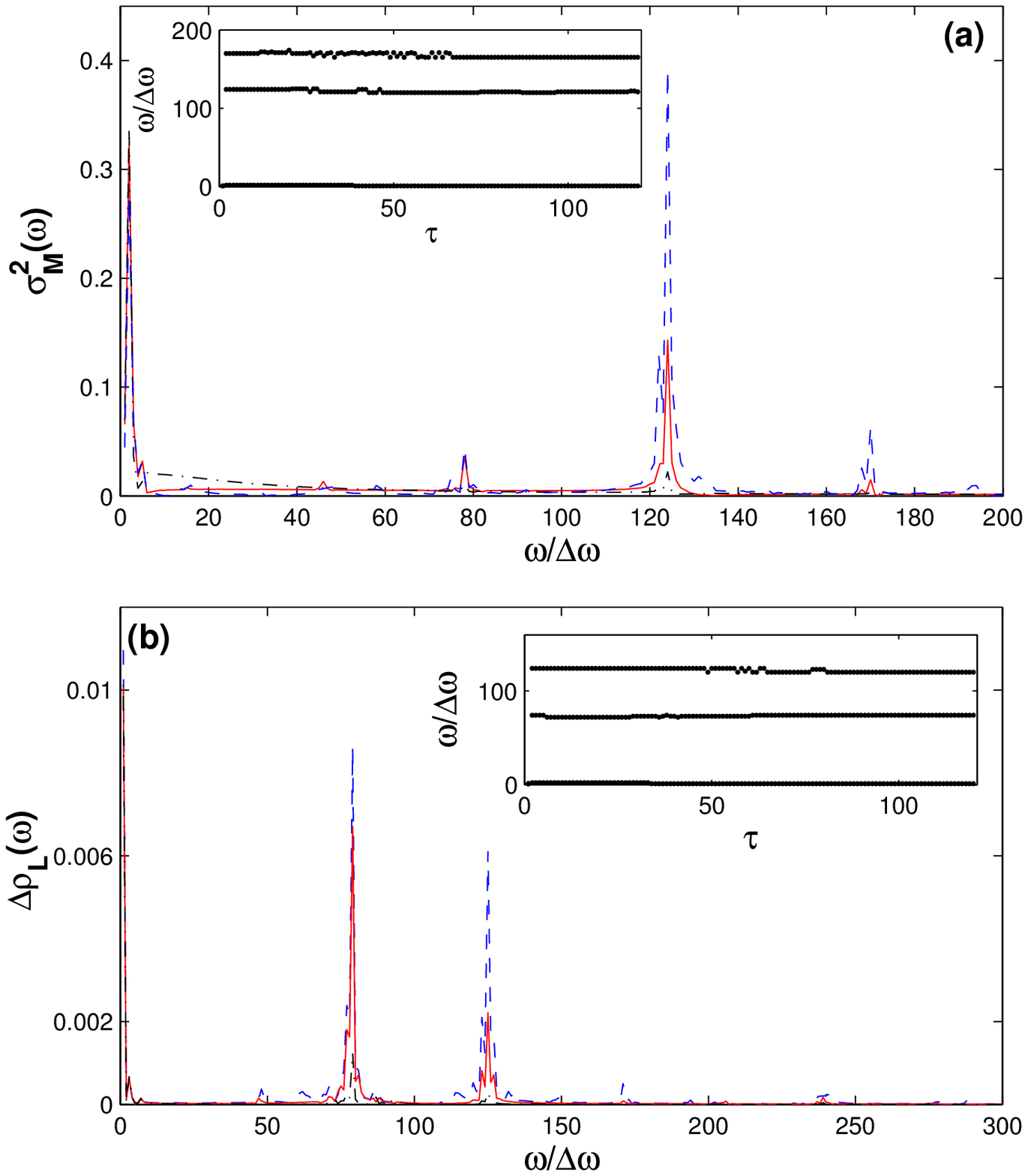}
              %  \vspace{1.0cm}
                \caption{(Color online) 
                (a) Frequency spectrum of the variance $\sigma _M^2(\omega )$ for time-dependent quenches of
                 the form of equation (11) with final interaction $g_{f}=0.1$ and rates $\tau=0\tau_{H}$ (blue dashed), $\tau=0.8\tau_{H}$ (red solid) and $\tau=3.0\tau_{H}$ (black dashed-dotted). The inset shows each branch
                 of the local breathing mode as a function of the quench rate $\tau$ (we incorporate 160 different rates in the
                range $0<\tau<120$). In (b) we present the spectrum of the intrawell asymmetry $\Delta\rho_{L}(\omega)$ for a time-dependent scenario with
                quench amplitude $\delta{g}=-4.90$ and
                rates $\tau=0\tau_{H}$ (blue dashed), $\tau=0.8\tau_{H}$ (red solid) and $\tau=3.0\tau_{H}$ (black dashed-dotted). The inset demonstrates the dependence of each branch of the cradle mode
                as a function of the quench rate $\tau$. Finally, note that we have used normalized frequency units $\omega/\Delta\omega$, with $\Delta\omega=2\pi/T$ and $T$ being
                the respective evolution time.}
\end{figure}

Finally, we study the effect of the finite ramping on the cradle
mode. Figure 9(b) presents the spectrum of the intrawell asymmetry
$\Delta {\rho _L}(\omega )$ for an abrupt quench in the
interparticle repulsion and two different quenches obeying the above
time-dependent law with different rates $\tau$ but same final
interaction as in the abrupt case. Moreover, in the inset we
demonstrate the evolution of each peak (without taking into account
its intensity) as a function of the ramp-rate $\tau$. We observe
that for larger rates $\tau$ the location of each frequency peak
remains essentially the same (inset) but the respective amplitude
tends to decrease, while for $\tau>9.0\tau_{H}$ the third peak that
refers to the second excited state in the left well has already been
eliminated. Increasing further the rate $\tau>30\tau_{H}$ one can
eliminate the cradle (intensity $\le10^{-5}$) approaching the
adiabatic region as also shown in Figure 7.

In the following section we turn to the study of the quench dynamics
induced by a modulation of the optical lattice depth examining its
dynamical response and the consequent excitation modes.

\subsection{Quench of the optical potential depth for filling $\nu<1$}

Here we consider a quench protocol which consists of a ramp-down of
the optical potential depth, thereby driving the system to a region
where the kinetic energy of the atoms dominates in comparison to the
potential energy. As we shall demonstrate, following this protocol
one can excite the cradle mode also for setups with filling $\nu<1$.
The system consists of five particles in an eight well setup but our
conclusions can be easily generalized for arbitrary filling factors.
To be self-consistent with the previous study we start from a
strongly interacting initial state with $g_{in}=5.0$, while the
lattice is assumed to be initially deep enough with a depth
$V_{0;in}=8.0$ to include the first three Wannier energy levels.
As usual, in order to interpret the dynamics induced by the quench
we should be aware of the characteristics of the initial ground
state. For a system with filling $\nu<1$ the one-body density
remains asymmetric even for strong interactions due to the low
population, while the delocalized fraction of particles permits the
presence of long range one particle correlations even in the
strongly repulsive regime \cite{Brouzos}.
\begin{figure}[t]
%\vspace{-15pt}
        \centering
                \includegraphics[width=1.00\columnwidth]{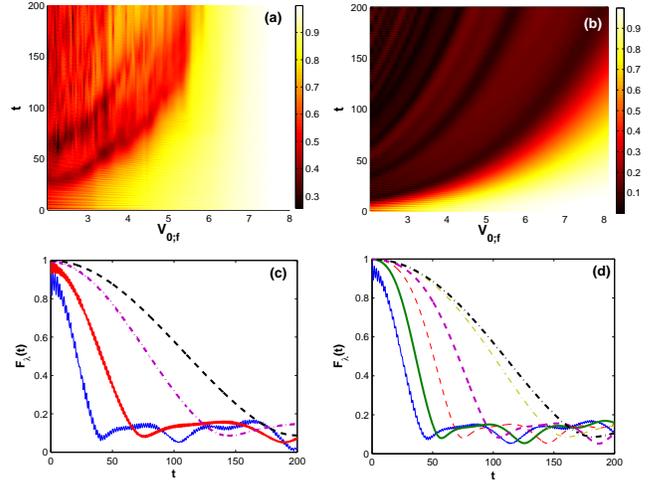}
              %  \vspace{1.0cm}
                \caption{(Color online) (a) Fidelity evolution as a function of different sudden negative quenches of the optical lattice depth. The
                system consists of five strongly interacting bosons ($g=5.0$) in an eight-well potential with $V_{0;in}=8.0$. (b) Same but
                for different quenches of the lattice depth and a simultaneous interaction quench to $g_{f}=0.02$.
                (c) Profiles of the fidelity evolution for
                different quench amplitudes $\delta{V_{0}}= -3.8$ (blue thin solid), $\delta{V_{0}}= -2.5$ (red thick solid),
                $\delta{V_{0}}= -0.9$ (magenta dashed-dotted), $\delta{V_{0}}= -0.2$ (black dashed) and a simultaneous interaction quench to $g_{f}=0.02$.
                In (d) we present profiles of the
                fidelity following a negative time-dependent quench of the potential depth to $V_{0;f}=4.0$ with different
                ramp-rates $\tau=0.4\tau_{H}$ (blue thin solid), $\tau=15.0\tau_{H}$ (green thick solid), $\tau=40.0\tau_{H}$ (red thin dashed), $\tau=100.0\tau_{H}$ (magenta thick dashed), $\tau=400.0\tau_{H}$ (yellow thin dashed-dotted)
                and $\tau=800.0\tau_{H}$ (black thick dashed-dotted) and a simultaneous interaction quench to $g_{f}=0.02$.}
\end{figure}

Let us firstly analyze the non-equilibrium
dynamics induced by a sudden ramp-down of the optical potential
depth at time $t$=0. The final Hamiltonian that governs the dynamics
following the above scenario is given by
\begin{equation}
\label{eq:16}H_{f}(g,{V_{0;f}}) = H_{0}(g,{V_{0;in}}) + \frac{{\delta
{V_0}}}{{{V_{0;in}}}}\sum\limits_{k = 1}^N {{V_{tr}}({x_k})},
\end{equation}
with $V_{0;in}$, $V_{0;f}$ being the initial and final potential
depth respectively, $\delta{V_{0}}=V_{0;f}-V_{0;in}<0$ due to the
reduction of the barrier and $V_{tr}$ being the lattice potential.
\begin{figure*}[t]
%\vspace{-15pt}
\centering
\includegraphics[width=0.80\textwidth]{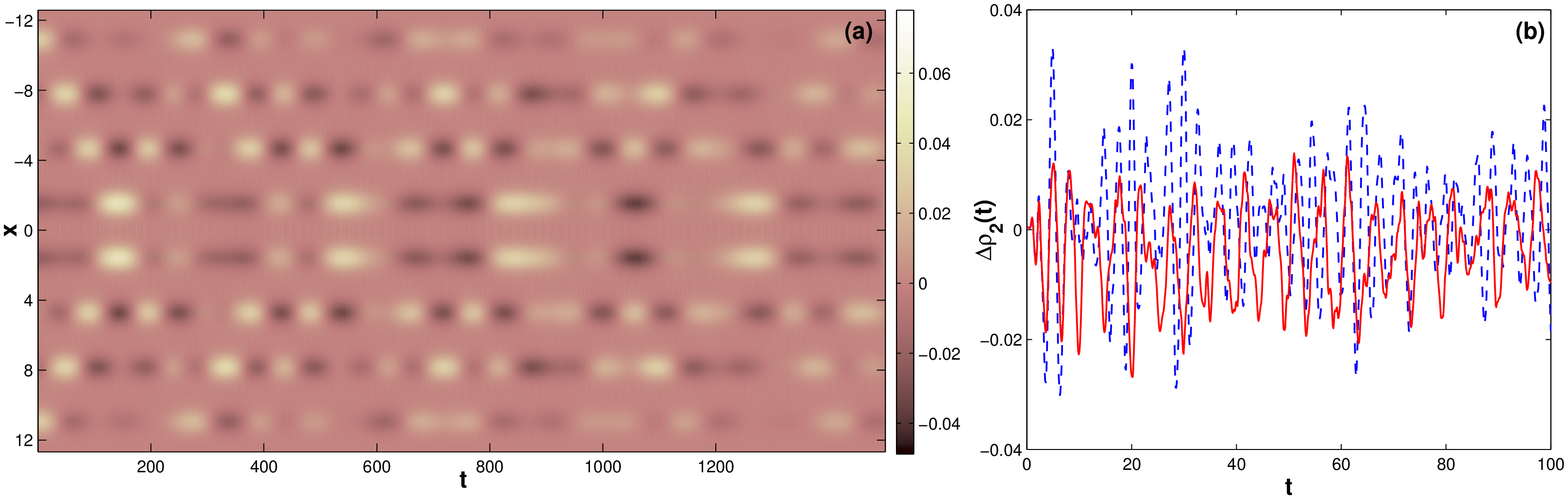}
%  \vspace{1.0cm}
\caption{(Color online) (a) The fluctuations $\delta\rho(x,t)$ of the one-body density, for an eight-well setup with $N=5$, caused by a sudden negative barrier quench from ${V_{0;in}} = 8.0$ 
to ${V_{0;f}} = 4.0$ and a simultaneous interaction quench from $g_{in}=5.0$ to $g_{f}=0.02$. 
%The initial state is the ground state of $N$=5 bosons confined in an eight-well setup with
%strong interparticle repulsion $g_{in}=5.0$ and depth ${V_{0;in}} = 8.0$. 
(b) The intrawell asymmetry $\Delta\rho_{2}(t)$ for the second well of the eight well setup for a barrier quench (red solid curve) to $V_{0;f}=4.0$ and for the combined quench scenario, i.e. barrier and interaction
quench, with final parameters $V_{0;f}=4.0$ and $g_{f}=0.02$ (blue dashed curve).}
\end{figure*}

To examine the response of the system after the quench we initially
rely on the fidelity ${F_\lambda }(t)$. We consider quenches of the
barrier down to $V_{0;f}=2.0$, where the lattice is quite shallow
and includes only the first Wannier energy level while the others
are considered as delocalized. Indeed, Figure 10(a) shows in a
transparent way the instantaneous fidelity as a function of the
final lattice depth. The rise of two different dynamical regions is
observed. In the first region ($5.0<V_{0;f}<8.0$) the overlap is
rather large with a minimum of the order of $80\%$, while in the
second region ($2.0<V_{0;f}<5.0$) it can even reach $25\%$ during
the evolution. As we shall demonstrate below, the response of the
system following this protocol is drastically different from that
obtained through an interaction quench for fillings $\nu<1$ where
the dynamics is dominated by the interwell tunneling. In particular,
one can excite more on-site dynamical modes and even use a barrier
quench on top of an interaction quench in order to excite the cradle
mode. To indicate the latter and also to trigger more efficiently
the dynamical modes from here on we mainly proceed by performing a
simultaneous barrier and an interaction quench to weak interactions,
i.e. $g_{f}=0.02$. Figure 10(b) presents the fidelity during the
dynamics induced by different quenches of the lattice depth and a
simultaneous interaction quench to $g_{f}=0.02$. The dynamical
response of the system shows four different regions during the
evolution. In the first (white part) the system is close to the
initial state with minimal percentage up to $80\%$, while the second
(yellow) and the third (light-red) regions indicate that the system
significantly departs from the initial state with a percentage of
the order of $50\%$ and $30\%$ respectively. The latter regions
correspond to transition states following the combined quench.
Finally, in the fourth section (dark-red) the system is driven to a
completely different state possessing a maximal overlap of the order
of $10\%$. In particular, for a fixed overlap a quadratic response
of the system as a function of the quench amplitude is observed.

To analyze further the response of the system, Figure 10(c)
illustrates some profiles of the fidelity in the course of the
dynamics for different quenches. The fidelity exhibits a quadratic
decay for short times, while after a characteristic time ${\tau
_{c}(\delta{V_{0}})}$ it oscillates around a constant value
$F_{0}(\delta{V_{0}})$, which depends on the quench amplitude such
that it is larger for a smaller quench. The observed short-time
quadratic behaviour can be easily explained as follows. Let
$\ket{\Psi_{0}}$ be the initial eigenstate of $H_{0}$, and
$\ket{\Psi{(\delta{t})}}$ the corresponding state after a short-time
interval $\delta{t}$. Then, the short-time expansion of the overlap
reads
\begin{equation}
\label{eq:17}{\left| {\left\langle {\Psi_0} \right|\left. {\Psi
(\delta{t})} \right\rangle }
\right|^2}=1-(\delta{t}/\tau_{c})^2+\mathcal{O}((\delta{t})^4),
\end{equation}
%where
$\tau_{c}^{-1}=(\Braket{\Psi(\delta{t})|H_f^2|\Psi(\delta{t})}-\Braket{\Psi(\delta{t})|H_{f}|\Psi(\delta{t})}^{2})^{1/2}$
is the quench characteristic time $\tau_{c}(\delta{V_{0}})$ or
so-called Zeno time \cite{Gorin,Peres}. Especially, we observe that
the time ${\tau _{c}(\delta{V_{0}})}$ depends on the quench
amplitude, i.e. for smaller quenches it becomes larger due to the
smaller energy difference between the pre and post-quench states,
and the system can not equilibrate fast. Furthermore, the rapid
small amplitude oscillations during the decay are a consequence of
the quantum interference and are predominantly due to the
overbarrier transport induced by the quench. Thus, they are also a
presignature of the cradle mode which is discussed below. The fact
that at least some frequencies of the cradle mode could be
indirectly observed in the fidelity spectrum is not surprising.
Indeed, from the expansion of the fidelity (see eq.(8)) in terms of
the number states it is obvious that when the contribution of the
excited band states, that refer to the cradle, is significant the
mode should also be observed in the fidelity spectrum. For smaller
quenches these amplitude oscillations fade out, thereby indicating
that the cradle is very weak.

Furthermore, in order to unravel the crossover between a diabatic
and an adiabatic quench, let us consider a time-dependent scenario
of the form $V(t;\tau) = {V_{0;in}} +
\left({V_{0;f}}-{V_{0;in}}\right)\tanh (t/\tau )$. To obtain the
physically relevant time-scales, let us rescale the time $t$ in
units of the quench characteristic time scale ${\tau _H}(\delta
V_{0}) \sim 1/\Delta \epsilon (\delta V_{0})$, where $\Delta
\epsilon  = \epsilon ({V_{0;in}};g_{in}) - \epsilon
({V_{0;f}};g_{f})$ is the energy difference between the pre-quenched
and the post-quenched system. Figure 10(d) demonstrates the fidelity
in the course of the dynamics for an interaction quench to
$g_{f}=0.02$ and the same final potential depth $V_{0;f}=4.4$
($V_{in}=8.0$) for different ramp rates $\tau$. A direct
observation, here, is that the system has a similar quadratic
response (for short-times) with respect to the quench, being
accompanied by small amplitude oscillations especially for fast
quenches, $\tau<28\tau_{H}$. This indicates that the characteristics
of the dynamics, such as the excitation modes, remain also for
finite rates. By considering large rates $\tau$ the switch-on
of the energy difference is sufficiently slow and an eigenstate of
the initial Hamiltonian becomes approximately an eigenstate for the
instantaneous final Hamiltonian. In this manner, we tend to approach
the adiabatic limit and the system equilibrates in a slower manner
while the small amplitude fast oscillations tend to disappear. Note
that for smaller relative quench amplitudes the adiabatic limit is
attained for smaller ramp-rates due to the reduced impact of the
quench.

The reduction of the lattice depth allows for an initially
delocalized boson to overcome the barrier if its kinetic energy
dominates with respect to the potential energy. Then, it is possible
to perform a collision with a second particle on the neighboring
site and a cradle state can be generated. Figure 11(a) illustrates from
the perspective of the relative density $\delta\rho(x,t)$ the
evolution of the system after a negative change of the lattice depth
from $V_{0;in}=8.0$ to $V_{0;f}=4.0$ and a simultaneous interaction
quench from $g_{in}=5.0$ to $g_{f}=0.02$. The dynamics shows the
propagation of interwell tunneling via the population transport
along the lattice, following the evolution of the bright regions.
The corresponding propagation velocity is smaller for a smaller
quench of the barrier. Furthermore, locally we observe the emergence
of the cradle mode for the inner-well dynamics as a consequence of
the overbarrier transport. However, this mode is hardly visible in
Figure 11(a) due to the presented long evolution time and possesses a
small amplitude as we shall demonstrate below.

The cradle mode refers to the inner-well oscillations between at
least two bosons in the same well. The dominant number states for
such a process exemplified using the first well are ${\left|
{2,0,1,1,1,0,...} \right\rangle _0}$ and ${\left| {2,0,1,1,1,0,...}
\right\rangle _1}$, with a straightforward extension for the
remaining wells. To identify the presence of the cradle mode for the
present setup we show in Figure 11(b) the intrawell asymmetry
($\Delta\rho_{\alpha}(t)$), for the second well of the lattice, as a
function of time, and two different quench scenarios, i.e. an
instantaneous ramp-down of the lattice depth (red solid curve) and
its combination with a simultaneous interaction quench to
$g_{f}=0.02$ (blue dashed curve). In the latter case the resulting
amplitude is larger, which is due to the simultaneous interaction
quench. For an incommensurate setup with filling $\nu>1$ this
amplitude is even more larger as the initially delocalized
particles, energetically close to the barrier, render the system
sensitive even to a small perturbation.

\section{Conclusions and Outlook}

We have investigated the quantum dynamics of strongly interacting
bosons following a quench either to a weakly interacting final state
or to a lattice with lowered barriers. The observed normal modes
consist of the interwell tunneling, a local breathing and a cradle
mode. Each of these modes have been explained in detail, among
others, within the concept of multiband Wannier number states which
capture the population of excited states. The dominant Fock space
for each mode has been identified showing the inclusion of
higher-band contributions. In this manner, conceptual differences
concerning the ingredients of each mode as well as the corresponding
excitation process in comparison with the case of positive
interaction quenched \cite{Mistakidis} superfluids have been
demonstrated.

The interwell tunneling refers mainly to a direct population
transport among the individual wells. On the
other hand, the local breathing mode refers to expansion and
contraction dynamics of the bosons in an individual well. The cradle
mode corresponds to a localized wavepacket oscillation. For the
interaction quench scenario where we start from a strongly
interacting state and quench back to weak-interactions it is shown
that the generation of the cradle mode is due to the initial
delocalization. Therefore it can be observed only for setups with
filling $\nu>1$, while for the case of $\nu\leq1$ it can be excited
only with the aid of a barrier quench. This is a major difference in
comparison to a positively interaction quenched superfluid where due
to import of energy in the system we allow for the over-barrier
transport independently of the filling factor. The fidelity function
has been employed in order to investigate the response of the system
and its long time evolution with respect to the quench amplitude, as
well as to show the dynamical crossover from a sudden to an
adiabatic parameter change. By considering time-dependent quenches,
i.e. different quench rates, or the modulation of various potential
parameters of the Hamiltonian we proposed scenarios to control the
excited modes by manipulating their frequencies.

Our developed understanding of the excitation modes as well as the
tunneling dynamics may pave the way to a control of the
nonequilibrium dynamics of such strongly correlated systems. For
instance, the finite ramp-rate of a time-dependent quench may allow
for the control of the normal modes or the transport of a definite
number of atoms. There are many ways to proceed in this direction.
As an example we mention the non-equilibrium dynamics of mixtures of
different bosonic species in order to unravel their excitation modes
or to create schemes for selective transport of an individual
bosonic component.

\section*{Appendix: The Computational Method, ML-MCTDHB}

Our analysis has been performed via the Multi-Layer Multi
Configuration Time-Dependent Hartree method for Bosons (ML-MCTDHB)
\cite{Cao,Kronke} which constitutes an ab-initio method for the
stationary properties but in particular the non-equilibrium quantum
dynamics of bosonic systems. For a single species it is identical to
MCTDHB which has been established \cite{Alon,Alon1,Streltsov} and
applied extensively \cite{Streltsov,Streltsov1,Alon2,Alon3}.

The advantage of the MCTDH-type methods \cite{Beck} in comparison to other exact
computational methods is the representation of the wavefunction by a
set of variationally optimal time-dependent orbitals. In turn, this
implies the truncation of the total Hilbert space to an optimal one
by employing a time-dependent moving basis in which the system can
be instantaneously optimally represented by time-dependent Hartree
products. The use of time-dependent orbitals is the key for the
numerically exact treatment, i.e. we need a much smaller set of
time-adaptive orbitals in order to achieve the same level of
accuracy compared to the case of a static basis. To be self
contained let us briefly introduce the basic concepts of the method
and discuss how it can be adapted to our purposes.

The main underlying idea of the MCTDHB method is to
solve the time-dependent Schr\"{o}dinger equation $\left( {i\hbar
{\partial _t} - H} \right)\Psi (x,t) = 0$ as an initial value
problem. The expansion of the many-body wavefunction which is a linear combination of time-dependent
permanents reads
\begin{equation}
\label{eq:A1}\left| {\Psi (t)} \right\rangle  = \sum\limits_{\vec n
} {{C_{\vec n }}(t)\left| {{n_1},{n_2},...,{n_M};t} \right\rangle },
\end{equation}
where $M$ is the number of orbitals and the summation is over all
possible combinations which retain the total number of bosons. The
permanents in terms of the creation operators $a_j^\dag (t)$ for the
$j - th$ orbital ${\varphi _j}(t)$ are given by
\begin{equation}
\begin{split}
\label{eq:A2}\left| {{n_1},{n_2},...,{n_M};t} \right\rangle =
\frac{1}{{\sqrt {{n_1}!{n_2}!...{n_M}!} }}{\left( {a_1^\dag }
\right)^{{n_1}}}{\left( {a_2^\dag }
\right)^{{n_2}}}\\\times...{\left( {a_M^\dag }
\right)^{{n_M}}}\left| {vac} \right\rangle,
\end{split}
\end{equation}
which satisfy the standard bosonic commutation relations $\left[
{{a_i}(t),{a_j}(t)} \right] = {\delta _{ij}}$, etc. To proceed
further, i.e. to determine the time-dependent wave function $\left|
\Psi \right\rangle$ we have to find the equations of motion for the
coefficients ${{C_{\vec n }}(t)}$ and the orbitals (which are both
time-dependent). For that purpose one can employ various schemes
such as the Lagrangian, McLachlan \cite{McLachlan} or the
Dirac-Frenkel \cite{Frenkel,Dirac} variational principle. Following
the Dirac-Frenkel variational principle ${\bra{\delta
\Psi}}{i{\partial _t} - \hat{ H}\ket{\Psi }}=0$ we can determine the
time evolution of all the coefficients ${{C_{\vec n }}(t)}$ in the
ansatz (14) and the time dependence for the orbitals $\left|
{{\varphi _j}} \right\rangle $. In this manner, we end up with a set
of $M$ non-linear integrodifferential equations of motion for the
orbitals which are coupled to the $\frac{(N+M-1)!}{N!(M-1)!}$ linear
equations of motion for the coefficients. These equations are the
well-known MCTDHB equations of motion
\cite{Alon,Streltsov,Alon1,Broeckhove}.

In terms of our implementation we have used a discrete variable
representation for the orbitals and a sin-DVR which intrinsically
introduces hard-wall boundaries at both ends of the potential (i.e.
zero value of the wave function on the first and the last grid
point). For the preparation of our initial state we rely on the
so-called relaxation method in terms of which we can obtain the
lowest eigenstates of the corresponding $n$-well setup. The key idea
is to propagate some initial wave function ${\Psi ^{(0)}}$ by the
non-unitary ${e^{ - H\tau }}$ (propagation in imaginary time). As
$\tau  \to \infty $, this exponentially damps out any contribution
but that stemming from the ground state like ${e^{ - {E_m}\tau }}$.
In turn, we change either the initial interparticle interaction or
the depth of the optical lattice abruptly or in a time-dependent
manner in order to study the evolution of $\Psi
({x_1},{x_2},..,{x_N};t)$ in the $n$-well potential within MCTDHB.
Finally, note that in order to ensure the convergence of our
simulations, e.g for the triple well, we have used up to 11 single
particle functions thereby observing a systematic convergence of our
results for sufficiently large spatial grids. Another criterion for
ensuring convergence is the population of the lowest occupied
natural orbital which is kept for each case below $0.1\%$.

\section*{Acknowledgments}
S.M. would like to thank P. Giannakeas, C. Morfonios and M. Mark for
fruitful discussions. S.M also thanks the Hamburgisches Gesetz zur F{\"o}rderung
des wissenschaftlichen und k{\"u}nstlerischen Nachwuchses (HmbNFG)
for a PhD Scholarship. L.C. and P.S gratefully acknowledge funding
by the Deutsche Forschungsgemeinschaft (DFG) in the framework of the
SFB 925 Light induced dynamics and control of correlated quantum
systems.

{}


\begin{thebibliography}{60}

\bibitem{Bloch}I. Bloch, J. Dalibard, and W. Zwerger,  Rev. Mod. Phys., \textbf{80}, 885 (2008).

\bibitem{Polkovnikov}A. Polkovnikov, K. Sengupta, A. Silva, and M. Vengalattore,  Rev. Mod. Phys., \textbf{83}, 863 (2011).

\bibitem{Weidemuller}M. Weidem{\"u}ller, A. Hemmerich, A. G{\"o}rlitz,
T. Esslinger, and T. W. H{\"a}nsch, Phys. Rev. Lett.,
\textbf{75}, 4583 (1995).

\bibitem{Hung}C. L. Hung, X. Zhang, L. C. Ha, S. K. Tung, N. Gemelke, and
C. Chin, New J. Phys., \textbf{13}, 075019 (2011).

\bibitem{Ronzheimer}J. P. Ronzheimer, M. Schreiber, S. Braun, S. S. Hodgman, S. Langer, I. P. McCulloch, F. Heidrich-Meisner, I. Bloch, and U. Schneider,
Phys. Rev. Lett., \textbf{110}, 205301 (2013).

\bibitem{Rigol}M. Rigol, V. Dunjko, and M. Olshanii, Nature, \textbf{452}, 854-858 (2008).

\bibitem{Cheneau}M. Cheneau, P. Barmettler, D. Poletti, M. Endres, P. Schauß, T. Fukuhara, C. Gross, I. Bloch, C. Kollath, and
S. Kuhr, Nature, \textbf{481} no. 7382  484-487 (2012).

\bibitem{Natu}S.S. Natu, and E. J. Mueller, Phys. Rev. A, \textbf{87}, 053607 (2013).

\bibitem{Altman}E. Altman, and A. Auerbach, Phys. Rev. Lett., \textbf{89}, 250404 (2002).

\bibitem{Chen}D. Chen, M. White, C. Borries, and B. DeMarco, Phys. Rev. Lett., \textbf{106}, 235304 (2011).

\bibitem{Haller}H. Elmar, R. Hart, M. J. Mark, J. G. Danzl, L. Reichs{\"o}llner, M. Gustavsson, M. Dalmonte, G. Pupillo,
and H.-C. N{\"a}gerl, Nature, \textbf{466} no. 7306: 597-600 (2010).

\bibitem{Mahmud}K. W. Mahmud, L. Jiang, P. R. Johnson, and E. Tiesinga, New J. Phys. \textbf{16} 103009 (2014).

\bibitem{Campbell} S. Campbell, M. A. García-March, T. Fogarty, and T. Busch, Phys. Rev. A \textbf{90}, 013617 (2014).

\bibitem{Mistakidis}S. I. Mistakidis, L. Cao, and P. Schmelcher, J. Phys. B: At. Mol. Opt. Phys. \textbf{47} 225303 (2014).

\bibitem{Kohn} W. Kohn, Phys. Rev. \textbf{123}, 1242 (1961).

\bibitem{Bonitz}M. Bonitz, K. Balzer, and R. Van Leeuwen, Phys. Rev. B, \textbf{76}, 045341 (2007).

\bibitem{Abraham}J. W. Abraham, and M. Bonitz, Contributions to Plasma Physics, \textbf{54}, 27-99 (2014).

\bibitem{Bauch1}S. Bauch, K. Balzer, C. Henning, and M. Bonitz, Phys. Rev. B, \textbf{80}, 054515 (2009).

\bibitem{Bauch}S. Bauch, D. Hochstuhl, K. Balzer, and M. Bonitz,
In Journal of Physics: Conference Series (Vol. \textbf{220}, No. 1, p.
012013). IOP Publishing (2010).

\bibitem{Abraham1}J. W. Abraham, K. Balzer, D. Hochstuhl, and M. Bonitz,
Phys. Rev. B, \textbf{86}, 125112 (2012).

\bibitem{Schmitz}R. Schmitz, S. Kr{\"o}nke, L. Cao, and P. Schmelcher, Phys. Rev. A, \textbf{88}, 043601 (2013).

\bibitem{Peotta}S. Peotta, D. Rossini, M. Polini, F. Minardi, and R. Fazio,
 Phys. Rev. Lett., \textbf{110}, 015302 (2013).

\bibitem{Makotyn}P. Makotyn, C. E. Klauss, D. L. Goldberger, E. A. Cornell, and  D. S. Jin, Nature Phys., \textbf{9}, 512 (2014).

\bibitem{Cao} L. Cao, S. Kr{\"o}nke, O. Vendrell, and P. Schmelcher,  J. Chem. Phys., \textbf{139}, 134103 (2013).

\bibitem{Kronke}S. Kr{\"o}nke, L. Cao, O. Vendrell, and P. Schmelcher, New J. Phys., \textbf{15}, 063018 (2013).

\bibitem{Alon} O. E. Alon, A. I. Streltsov, and L. S. Cederbaum,  J. Chem. Phys., \textbf{127}, 154103 (2007).

\bibitem{Alon1}O. E. Alon, A. I. Streltsov, and L. S. Cederbaum, Phys. Rev. A, \textbf{77}, 033613 (2008).

\bibitem{Olshanii}M. Olshanii, Phys. Rev. Lett., \textbf{81}, 938 (1998).

\bibitem{Grimm}R. Grimm, M. Weidem{\"u}ller, and Y. B. Ovchinnikov,  Adv. At. Mol. Opt. Phys., \textbf{42}, 95-170 (2000).

\bibitem{Duine}R. A. Duine, and H. T. Stoof,  Phys. Rep., \textbf{396}, 115-195 (2004).

\bibitem{Chin}C. Chin, R. Grimm, P. Julienne, and E. Tiesinga,  Rev. Mod. Phys., \textbf{82}, 1225 (2010).

\bibitem{Kim}J. I. Kim, V. S. Melezhik, and P. Schmelcher, Phys. Rev. Lett., \textbf{97}, 193203 (2006).

\bibitem{Giannakeas}P. Giannakeas, F. K. Diakonos, and P. Schmelcher,  Phys. Rev. A, \textbf{86}, 042703 (2012).

\bibitem{Gorin}T. Gorin, T. Prosen, T. H. Seligman, and M. \v{Z}nidari\v{c}, Phys. Rep., \textbf{435}, 33-156 (2006).

\bibitem{Spekkens}R. W. Spekkens, and J. E. Sipe, Phys. Rev. A, \textbf{59}, 3868 (1999).

\bibitem{Klaiman}S. Klaiman, N. Moiseyev, and L. S. Cederbaum, Phys. Rev. A, \textbf{73}, 013622 (2006).

\bibitem{Mueller}E. J. Mueller, T. L. Ho, M. Ueda, and G. Baym, Phys. Rev. A, \textbf{74}, 33612 (2006).

\bibitem{Penrose}O. Penrose, and L. Onsager, Phys. Rev., \textbf{104}, 576 (1956).

\bibitem{Fisher}M. P. Fisher,  P. B. Weichman, G. Grinstein, and D. S. Fisher,
 Phys. Rev. B, \textbf{40}, 546 (1989).

\bibitem{Freericks}J. K. Freericks, and H. Monien, EPL, \textbf{26}, 545 (1994).

\bibitem{Freericks1}J. K. Freericks, and H. Monien, Phys. Rev. B, \textbf{53}, 2691 (1996).

\bibitem{Sakmann}K. Sakmann, A. I. Streltsov, O. E. Alon, and L. S. Cederbaum, Phys. Rev. A, \textbf{89}, 23602 (2014).

\bibitem{Bravyi}S. Bravyi, M. B. Hastings, and F. Verstraete, Phys. Rev. Lett., \textbf{97}, 050401 (2006).

\bibitem{Sakmann1}K. Sakmann,  A. I. Streltsov, O. E. Alon, , and L. S. Cederbaum, Phys. Rev. A, \textbf{78}, 023615 (2008).

\bibitem{Brouzos} I. Brouzos, S. Z{\"o}llner, and P. Schmelcher, Phys. Rev. A, {\bf 81}, 053613 (2010).

\bibitem{Peres}A. Peres, Am. J. Phys., \textbf{48}, 913 (1980).

\bibitem{Streltsov} A. I. Streltsov, O. E. Alon, and L. S. Cederbaum, Phys. Rev. Lett., \textbf{99}, 030402 (2007).

\bibitem{Streltsov1} A. I. Streltsov, K. Sakmann, O. E. Alon, and L. S. Cederbaum, Phys. Rev. A, \textbf{83}, 043604 (2011).

\bibitem{Alon2} O. E. Alon, A. I. Streltsov, and L. S. Cederbaum, Phys. Rev. A, \textbf{76}, 013611 (2007).

\bibitem{Alon3} O. E. Alon, A. I. Streltsov, and L. S. Cederbaum, Phys. Rev. A, \textbf{79}, 022503 (2009).

\bibitem{Beck}M. H. Beck, A. J{\"a}ckle, G. A. Worth, and H. D. Meyer, Phys. Rep.,
\textbf{324} (1999).

\bibitem{McLachlan} A. D. McLachlan,  Mol. Phys., \textbf{8}, 39-44 (1964).

\bibitem{Frenkel} J. Frenkel, Wave mechanics, (pp.
423-28). Oxford (1934).

\bibitem{Dirac} P. A. Dirac, (1930, July). Proc. Camb. Phil. Soc.
(Vol. \textbf{26}, No. 03, pp. 376-385). Cambridge University Press.

\bibitem{Broeckhove} J. Broeckhove, L. Lathouwers, E. Kesteloot, and Van Leuven,
P. Chem. Phys. Lett., \textbf{149}, 547-550 (1988).


\end{thebibliography}
\end{document}